\documentclass[conference]{IEEEtran}
\IEEEoverridecommandlockouts
\usepackage{amsmath,amssymb,amsfonts}
\usepackage{algorithmic}
\usepackage{graphicx}
\usepackage{textcomp}
\usepackage{xcolor}

\usepackage[style=ieee]{biblatex}
\usepackage{software-biblatex}

\usepackage{tikz}
\usepackage{pgfplots}
\usepackage{pgfmath}
\pgfplotsset{compat=1.17}

\usepackage{subcaption}
\usepackage{multirow}

\usetikzlibrary{patterns}

\definecolor{RYB1}{RGB}{166,206,227}  
\definecolor{RYB2}{RGB}{31,120,180}   
\definecolor{RYB3}{RGB}{178,223,138}  
\definecolor{RYB4}{RGB}{51,160,44}    
\definecolor{RYB5}{RGB}{251,154,153}  
\definecolor{RYB6}{RGB}{227,26,28}    
\definecolor{RYB7}{RGB}{253,191,111}  
\definecolor{RYB8}{RGB}{255,127,0}    
\definecolor{RYB9}{RGB}{202,178,214}  
\definecolor{RYB10}{RGB}{160,60,140}  
\definecolor{RYB11}{RGB}{255,255,153} 
\definecolor{RYB12}{RGB}{177,89,40}   

\definecolor{code}{rgb}{0, 0, 0}

\pgfplotscreateplotcyclelist{will}{%
RYB2!50!black,every mark/.append style={fill=RYB2},mark=*,mark repeat=5\\%
RYB4!50!black,every mark/.append style={fill=RYB4},mark=*,mark repeat=5\\%
RYB6!50!black,every mark/.append style={fill=RYB6},mark=*,mark repeat=5\\%
RYB8!50!black,every mark/.append style={fill=RYB8},mark=*,mark repeat=5\\%
RYB10!50!black,every mark/.append style={fill=RYB10},mark=*,mark repeat=5\\%
RYB12!50!black,every mark/.append style={fill=RYB12},mark=*,mark repeat=5\\%
black,every mark/.append style={fill=white!50!black},mark=*,mark repeat=5\\%
RYB2!50!black,every mark/.append style={fill=RYB2},mark=*,mark repeat=5\\%
RYB4!50!black,densely dashed,every mark/.append style={fill=RYB4},mark=*,mark repeat=5\\%
RYB6!50!black,densely dashed,every mark/.append style={fill=RYB6},mark=*,mark repeat=5\\%
RYB8!50!black,densely dashed,every mark/.append style={fill=RYB8},mark=*,mark repeat=5\\%
RYB10!50!black,densely dashed,every mark/.append style={fill=RYB10},mark=*,mark repeat=5\\%
RYB12!50!black,densely dashed,every mark/.append style={fill=RYB12},mark=*,mark repeat=5\\%
black,densely dashed,every mark/.append style={fill=white!50!black},mark=*,mark repeat=5\\%
RYB2!50!black,densely dashed,every mark/.append style={fill=RYB2},mark=*,mark repeat=5\\%
}

\newcommand{\NB}{N\!B}

\begin{document}

\title{Optimizing High-Performance Linpack for Exascale Accelerated Architectures}

\author{
\IEEEauthorblockN{Noel Chalmers}
\IEEEauthorblockA{\textit{Advanced Micro Devices Inc.} \\
Austin, Texas, USA \\
noel.chalmers@amd.com}
\and
\IEEEauthorblockN{Jakub Kurzak}
\IEEEauthorblockA{\textit{Advanced Micro Devices Inc.} \\
Oak Ridge, Tennessee, USA \\
jakub.kurzak@amd.com}
\and
\IEEEauthorblockN{Damon McDougall}
\IEEEauthorblockA{\textit{Advanced Micro Devices Inc.} \\
Austin, Texas, USA \\
damon.mcdougall@amd.com}
\and
\IEEEauthorblockN{Paul T. Bauman}
\IEEEauthorblockA{\textit{Advanced Micro Devices Inc.} \\
Austin, Texas, USA \\
paul.bauman@amd.com}
}
\maketitle

\begin{abstract}
We detail the performance optimizations made in rocHPL, AMD's open-source implementation of the High-Performance Linpack (HPL) benchmark targeting accelerated node architectures designed for exascale systems such as the Frontier supercomputer. The implementation leverages the high-throughput GPU accelerators on the node via highly optimized linear algebra libraries, as well as the entire CPU socket to perform latency-sensitive factorization phases. We detail novel performance improvements such as a multi-threaded approach to computing the panel factorization phase on the CPU, time-sharing of CPU cores between processes on the node, as well as several optimizations which hide MPI communication. We present some performance results of this implementation of the HPL benchmark on a single node of the Frontier early access cluster at Oak Ridge National Laboratory, as well as scaling to multiple nodes.
\end{abstract}


\section{Introduction}
In June of 2022, the Frontier supercomputer housed at Oak Ridge National Laboratory (ORNL) debuted on the Top500 list of supercomputers in the top position with a High-Performance Linpack (HPL) \cite{dongarra2003linpack} score of 1.1 ExaFLOPS \cite{top500}. Over twice the score of the previous top supercomputer, Frontier was the first supercomputer ever to achieve over one ExaFLOPS in HPL, marking it as the first true exascale machine. Shortly afterwards, AMD's optimized implementation of HPL, named \texttt{rocHPL} \cite{rochpl}, was made open-source and freely available. A variant of this HPL implementation, with optimized communication performance provided by Hewlett Packard Enterprise (HPE), was run on Frontier to achieve over one ExaFLOPS of overall performance. In this paper, we detail some of the performance optimizations which helped achieve this score with the expectation that these performance optimizations provides useful information for users optimizing HPL on heterogeneous architectures. 

HPL is one of many benchmarks designed to measure some aspects of a computer system. Other common benchmarks include the High-Performance Conjugate Gradients (HPCG) benchmark which stresses the system's main memory bandwidth and system-wide all-reduce performance, and the High Performance Linpack - Mixed Precision (HPL-MxP) benchmark which stresses the system's computational throughput of mixed- and lower-precision math operations. As with these other benchmarks, HPL effectively stresses several aspects of a computer system including the 64-bit floating-point computation rate, network bandwidth, and network latency for extended periods of time while drawing essentially the peak amount of power the system can use. This makes HPL an incredibly useful stress test for validating a new computer system's reliability and overall performance.

The high FLOP rate in HPL on Frontier is owed almost entirely to its GPU-accelerated node architecture and high-speed network. The presence of accelerators is a growing trend in high-performance computing. Indeed, as of June 2022 seven of the top ten supercomputers on the Top500 list have GPU-accelerated node architectures. In Frontier's case, each node is comprised of a single 64-core AMD EPYC CPU, four AMD Instinct MI250X GPU accelerators. The EPYC CPU and MI250X GPUs all leverage AMD's advanced packaging as Multi-Chip Modules (MCMs). The CPU socket is comprised of eight 8-core Core Complex Dies (CCDs) and an IO die, while each MI250X GPU is comprised of two Graphics Compute Dies (GCDs). The GCDs are connected to one another, and to the CPU socket, via AMD Infinity Fabric. With this architecture, the MI250X accelerators in each Frontier node contribute over 98\% of the node's peak FLOPS rate.

The computation performed in the HPL benchmark is the solution of an $N\times N$ random linear system of equations with a blocked Gaussian elimination method with partial pivoting. The $N \times N$ matrix, $A$, is distributed into a $P\times Q$ MPI process grid via a 2D block-cyclic distribution for load balancing. Leveraging the high FLOP rate of the GPU accelerators in HPL requires careful consideration of the implementation of its four main phases, each with different computational character. These phases are: 
\begin{itemize}
\item Panel factorization (FACT) - The leading $\NB$ columns of $A$ are $LU$ factorized. This involves $\NB$ small collectives among the $P$ processes performing the factorization in order to find the pivot rows.
\item Panel broadcast (LBCAST) - The trailing matrix below the current $\NB \times \NB$ diagonal block is broadcast to all other processes in the grid. 
\item Row-swapping (RS) - The $\NB$ pivots determined during FACT are applied to the remaining columns of $A$
\item Trailing update (UPDATE) - A rank $\NB$ update is applied to the trailing submatrix of $A$. This phase consists of a computationally demanding triangular inverse, a.k.a. DTRSM, and a matrix-matrix multiplication routine, a.k.a. DGEMM.
\end{itemize}
Classically, the vast majority of time in the HPL benchmark is spent inside DGEMM routines inside the trailing update phase. These routines are often highly optimized on different hardware which enables HPL scores to achieve significant fractions of the hardware's peak floating point computation rate. 

As accelerators became more prevalent over the past decade, there have been several works which studied leveraging them in HPL. A natural approach is to use accelerators to improve the performance of the large DGEMM computations in HPL, keeping the matrix in the CPU memory and offloading the DGEMM operations to the accelerator by dividing it into smaller pieces and processing piece by piece, interleaved with data transfers. Endo \& Matsuoka \cite{endo2008massive} first studied using ClearSpeed SIMD accelerators to improve DGEMM performance in HPL in a heterogeneous cluster where only some nodes contained accelerators. This idea of offloading the DGEMM work to accelerators was also applied for HPL on the Roadrunner supercomputer by Kistler et al. \cite{kistler2009petascale} which used the IBM PowerXCell 8i accelerators. Fatica \cite{fatica2009accelerating} described this approach for GPGPUs using CUDA, wherein they describe a pipe-lining strategy for moving sections of the input/output matrices to/from the GPU to accelerate both DGEMM and DTRSM routines and hiding this data motion behind the computation time on the GPGPU. Demonstrations of scaling of this pipe-lining strategy to full clusters were described by Wang et al. \cite{wang2011optimizing} and Rohr et al. \cite{rohr2011multi}. Other authors considered similar approaches for other programming models such as OpenCL \cite{jo2014accelerating}, and other accelerators such as the Intel Xeon Phi \cite{heinecke2013design}, and even clusters which mixed different types of accelerators \cite{endo2010linpack}.

In more recent works, several authors have noted the increases in computation rates of accelerators have out-paced the bandwidth improvements in the host-accelerator links. Indeed, some modern GPU accelerators, including the MI250X GPUs, also include specialized hardware units which further accelerate the compute rate of matrix-matrix multiplications. This has made the pipe-lining strategy described by Fatica \cite{fatica2009accelerating} for accelerating DGEMMs and other routines in HPL impractical. In order to hide the data motion between host and accelerator, the amount of computation done in each kernel must be dramatically increased, usually leading to unreasonably large blocking parameters in HPL which induces bottlenecks in other phases. To alleviate this, Tan et al. \cite{tan2021optimizing} and Kim et al. \cite{kim2022snuhpl} present HPL implementations on modern GPU accelerators where the entire problem is stored in the GPUs' memory, rather than host DDR, an idea that originally appeared in Kistler et al. \cite{kistler2009petascale}. This has the benefit of removing the need to move data for large computational routines to/from the accelerator. Several large MPI communication phases can then also leverage GPU-aware MPI routines to move data directly between different GPUs' HBM and leverage fast hardware links between GPUs on-node when available. Some complications in this implementation arise, however. The panel factorization remains a complex communication- and latency-sensitive phase in HPL, which is not well-suited for fine-grain parallelism on accelerators. Furthermore, MPI communications must still be coordinated by the host process and overlapping these communications with useful work on the accelerator can be challenging.

Both Tan et al. \cite{tan2021optimizing} and Kim et al. \cite{kim2022snuhpl} opt to utilize the host CPU to perform the panel factorization, transferring only the needed data from/to the GPU at each iteration. This reduced amount of data motion allows for the FACT and LBCAST phases to easily overlap with trailing update computation on the GPU. Communication required to perform the row-swapping phase then introduces GPU idle time, which both studies resolve by splitting the row-swapping and trailing update into several smaller pieces and pipelining to hide communication by smaller trailing updates. Tan et al. implement even further pipelining using multiple CPU threads to advance the panel broadcast phase while panel factorization progresses. This approach, which potentially causes some congestion of the network interfaces, is not used by Kim et al. who instead use NVIDIA's NCCL communication library for GPU-direct communication which uses GPU kernels for data motion, making overlapping with other communication impractical. 

\begin{figure}[tbp]
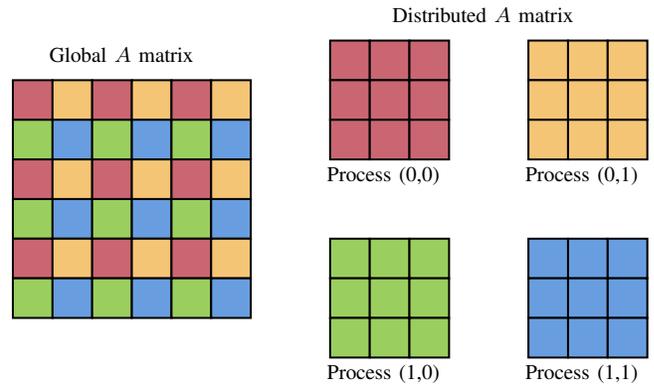

\centering
\input Figures/cyclic-distribution
\caption{2D block cyclic distribution of a matrix. The global $N\times N$ matrix is blocked into $\NB\times\NB$ panels, which are distributed among a $P\times Q$ grid of processes. Figure shows an example of a distribution into a $2\times 2$ process grid. }
\label{fig:cyclic-distribution}
\end{figure}


In this paper, we detail some of the optimizations we have implemented for HPL to improve performance on GPU-accelerated node architectures such as Frontier. In particular, we detail a multi-threading strategy for extracting data parallelism in the inherently serial panel factorization phase, overlapping CPU and GPU computation, and hiding GPU-GPU communication time via a split trailing update formulation. We then present some performance results of this implementation of the HPL benchmark on the Crusher cluster at ORNL showing the efficacy of our optimizations in hiding MPI communication time and demonstrating good scaling performance to multiple nodes, and afterwards give some concluding remarks. 

\begin{figure*}[tbp]
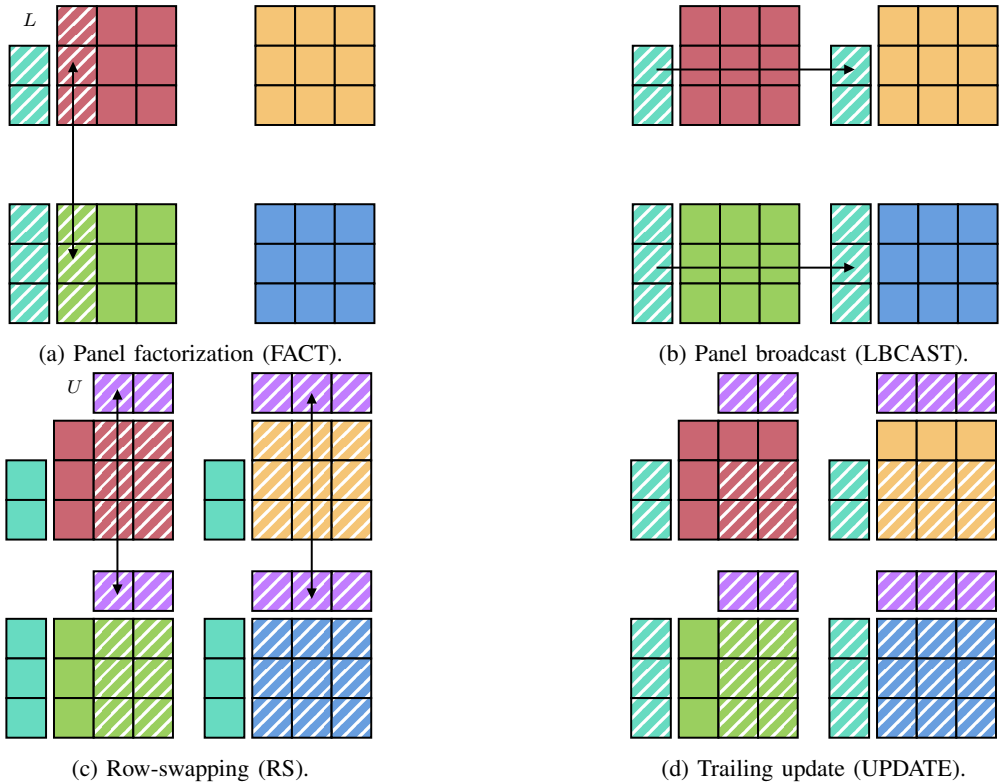

\centering
\begin{subfigure}[t]{0.45\textwidth}
    \centering
    \input Figures/pdfact
    \caption{Panel factorization (FACT).}
    \label{fig:pdfact}
\end{subfigure}
\begin{subfigure}[t]{0.45\textwidth}
    \centering
    \input Figures/lbcast
    \caption{Panel broadcast (LBCAST).}
    \label{fig:lbcast}
\end{subfigure}

\begin{subfigure}[t]{0.45\textwidth}
    \centering
    \input Figures/laswp
    \caption{Row-swapping (RS).}
    \label{fig:laswp}
\end{subfigure}
\begin{subfigure}[t]{0.45\textwidth}
    \centering
    \input Figures/update
    \caption{Trailing update (UPDATE).}
    \label{fig:update}
\end{subfigure}
\caption{The four main phases at each iteration of HPL's factorization algorithm. The figures show the phases on an example $2\times 2$ process grid. In each phase, we show with a patterned fill what panels of each process's local matrix are accessed. We also show with arrows what processes communicate in each phase.}
\label{fig:phases}
\end{figure*}

\section{HPL Overview}

The HPL benchmark begins by generating a distributed $N \times N$ double-precision matrix $A$ on a two-dimensional grid of $P\times Q$ processes. For load balancing, the global matrix $A$ is blocked into $\NB \times\NB$ sized panels, and these panels are distributed to the process grid in a 2D block-cyclic fashion. An example of this distribution is depicted graphically in Figure \ref{fig:cyclic-distribution} for a $2\times2$ process grid. A length $N$ right-hand-side vector $\mathbf{b}$ is also generated and appended to $A$ to form an $N\times(N+1)$ augmented system. 

The linear system is solved via a blocked Gaussian elimination algorithm, with partial pivoting. By treating $A$ as an augmented system, the linear system $A \mathbf{x} = \mathbf{b}$ is essentially transformed into the upper triangular system $U \mathbf{x} = \hat{\mathbf{b}} = L^{-1}P^{-1} \mathbf{b}$, where $A = PLU$ is the $LU$-factorization of $A$ with row-pivoting. After this transformation, the solution vector $\mathbf{x}$ is readily found by applying $U^{-1}$. 

\begin{figure*}[tbph]
\centering
\tikzset{every picture/.style={line width=0.75pt}} 

\begin{tikzpicture}[x=0.75pt,y=0.75pt,yscale=-1,xscale=1]

\draw  [fill={rgb, 255:red, 214; green, 214; blue, 214 }  ,fill opacity=1 ] (92,60) -- (660,60) -- (660,80) -- (92,80) -- cycle ;
\draw  [fill={rgb, 255:red, 214; green, 214; blue, 214 }  ,fill opacity=1 ] (92,100) -- (660,100) -- (660,120) -- (92,120) -- cycle ;
\draw  [fill={rgb, 255:red, 214; green, 214; blue, 214 }  ,fill opacity=1 ] (92,140) -- (660,140) -- (660,160) -- (92,160) -- cycle ;
\draw  [fill={rgb, 255:red, 214; green, 214; blue, 214 }  ,fill opacity=1 ] (92,20) -- (660,20) -- (660,40) -- (92,40) -- cycle ;
\draw  [fill={rgb, 255:red, 231; green, 135; blue, 135 }  ,fill opacity=1 ] (148,64) .. controls (148,61.79) and (149.79,60) .. (152,60) -- (216,60) .. controls (218.21,60) and (220,61.79) .. (220,64) -- (220,76) .. controls (220,78.21) and (218.21,80) .. (216,80) -- (152,80) .. controls (149.79,80) and (148,78.21) .. (148,76) -- cycle ;
\draw  [fill={rgb, 255:red, 201; green, 118; blue, 218 }  ,fill opacity=1 ] (568,141.6) .. controls (568,140.72) and (568.72,140) .. (569.6,140) -- (574.4,140) .. controls (575.28,140) and (576,140.72) .. (576,141.6) -- (576,158.4) .. controls (576,159.28) and (575.28,160) .. (574.4,160) -- (569.6,160) .. controls (568.72,160) and (568,159.28) .. (568,158.4) -- cycle ;
\draw  [fill={rgb, 255:red, 126; green, 231; blue, 204 }  ,fill opacity=1 ] (632,141.6) .. controls (632,140.72) and (632.72,140) .. (633.6,140) -- (638.4,140) .. controls (639.28,140) and (640,140.72) .. (640,141.6) -- (640,158.4) .. controls (640,159.28) and (639.28,160) .. (638.4,160) -- (633.6,160) .. controls (632.72,160) and (632,159.28) .. (632,158.4) -- cycle ;
\draw  [fill={rgb, 255:red, 154; green, 190; blue, 129 }  ,fill opacity=1 ] (112,143.2) .. controls (112,141.43) and (113.43,140) .. (115.2,140) -- (124.8,140) .. controls (126.57,140) and (128,141.43) .. (128,143.2) -- (128,156.8) .. controls (128,158.57) and (126.57,160) .. (124.8,160) -- (115.2,160) .. controls (113.43,160) and (112,158.57) .. (112,156.8) -- cycle ;
\draw  [fill={rgb, 255:red, 154; green, 190; blue, 129 }  ,fill opacity=1 ] (128,144) .. controls (128,141.79) and (129.79,140) .. (132,140) -- (564,140) .. controls (566.21,140) and (568,141.79) .. (568,144) -- (568,156) .. controls (568,158.21) and (566.21,160) .. (564,160) -- (132,160) .. controls (129.79,160) and (128,158.21) .. (128,156) -- cycle ;
\draw  [fill={rgb, 255:red, 148; green, 185; blue, 235 }  ,fill opacity=1 ] (576,24) .. controls (576,21.79) and (577.79,20) .. (580,20) -- (628,20) .. controls (630.21,20) and (632,21.79) .. (632,24) -- (632,36) .. controls (632,38.21) and (630.21,40) .. (628,40) -- (580,40) .. controls (577.79,40) and (576,38.21) .. (576,36) -- cycle ;
\draw  [fill={rgb, 255:red, 253; green, 246; blue, 139 }  ,fill opacity=1 ] (128,104) .. controls (128,101.79) and (129.79,100) .. (132,100) -- (144,100) .. controls (146.21,100) and (148,101.79) .. (148,104) -- (148,116) .. controls (148,118.21) and (146.21,120) .. (144,120) -- (132,120) .. controls (129.79,120) and (128,118.21) .. (128,116) -- cycle ;
\draw  [fill={rgb, 255:red, 253; green, 246; blue, 139 }  ,fill opacity=1 ] (220,104) .. controls (220,101.79) and (221.79,100) .. (224,100) -- (236,100) .. controls (238.21,100) and (240,101.79) .. (240,104) -- (240,116) .. controls (240,118.21) and (238.21,120) .. (236,120) -- (224,120) .. controls (221.79,120) and (220,118.21) .. (220,116) -- cycle ;
\draw  [fill={rgb, 255:red, 148; green, 185; blue, 235 }  ,fill opacity=1 ] (148,24) .. controls (148,21.79) and (149.79,20) .. (152,20) -- (216,20) .. controls (218.21,20) and (220,21.79) .. (220,24) -- (220,36) .. controls (220,38.21) and (218.21,40) .. (216,40) -- (152,40) .. controls (149.79,40) and (148,38.21) .. (148,36) -- cycle ;
\draw  [fill={rgb, 255:red, 148; green, 185; blue, 235 }  ,fill opacity=1 ] (240,24) .. controls (240,21.79) and (241.79,20) .. (244,20) -- (292,20) .. controls (294.21,20) and (296,21.79) .. (296,24) -- (296,36) .. controls (296,38.21) and (294.21,40) .. (292,40) -- (244,40) .. controls (241.79,40) and (240,38.21) .. (240,36) -- cycle ;
\draw  [dash pattern={on 4.5pt off 4.5pt}]  (112,8) -- (112,172) ;
\draw  [dash pattern={on 4.5pt off 4.5pt}]  (640,8) -- (640,172) ;
\draw   (136,176) -- (120,176) -- (120,164) ;
\draw   (556,192) -- (636,192) -- (636,164) ;
\draw   (556,176) -- (572,176) -- (572,164) ;

\draw (167,26) node [anchor=north west][inner sep=0.75pt]  [font=\footnotesize] [align=left] {FACT};
\draw (246,26) node [anchor=north west][inner sep=0.75pt]  [font=\footnotesize] [align=left] {LBCAST};
\draw (593,26) node [anchor=north west][inner sep=0.75pt]  [font=\footnotesize] [align=left] {RS};
\draw (167,66) node [anchor=north west][inner sep=0.75pt]  [font=\footnotesize] [align=left] {FACT};
\draw (309,146) node [anchor=north west][inner sep=0.75pt]  [font=\footnotesize] [align=left] {UPDATE};
\draw (57,26) node [anchor=north west][inner sep=0.75pt]  [font=\footnotesize] [align=left] {MPI};
\draw (52,65) node [anchor=north west][inner sep=0.75pt]  [font=\footnotesize] [align=left] {CPU};
\draw (33,105) node [anchor=north west][inner sep=0.75pt]  [font=\footnotesize] [align=left] {Transfer};
\draw (53,145) node [anchor=north west][inner sep=0.75pt]  [font=\footnotesize] [align=left] {GPU};
\draw (139,170) node [anchor=north west][inner sep=0.75pt]  [font=\footnotesize] [align=left] {Look-ahead UPDATE};
\draw (485,186) node [anchor=north west][inner sep=0.75pt]  [font=\footnotesize] [align=left] {Row Scatter};
\draw (485,170) node [anchor=north west][inner sep=0.75pt]  [font=\footnotesize] [align=left] {Row Gather};

\end{tikzpicture}
\caption{Diagram of the execution of a single iteration in HPL's factorization. Diagram shows utilization of both CPU and GPU, as well as the data transfer between them. MPI activity in different phases is also represented on the timeline. When the walltime of the UPDATE phase is large, we observe that the computation on the GPU can effectively hide all phases except for RS.}
\label{fig:timeline}
\end{figure*}
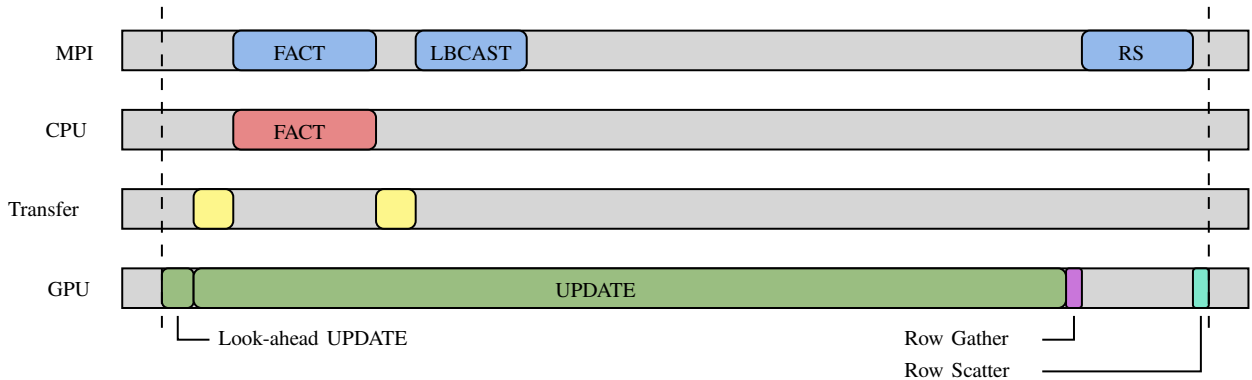

The blocked algorithm proceeds iteratively along the diagonal of $A$. Each iteration then consists of four main phases which themselves consist of varying levels of computation and communication between processes. To begin, at each iteration the block of $\NB$ columns at the current position along the diagonal of $A$ are $LU$ factored, applying row pivoting only within these $\NB$ columns and leaving the rest of the matrix unchanged. This phase, called the `panel factorization' (FACT) stage, is shown graphically in Figure \ref{fig:pdfact} for a $2\times2$ process grid. Only the leftmost block of panels is accessed, as shown in the patterned panels in the figure, and only in the processes which own sections of these $\NB$ columns. The processes participating in the panel factorization must frequently communicate in order to determine the $\NB$ row pivots to apply as the factorization progresses. Communication is indicated in the figure by arrows between these processes. Each communication to determine the distinct row to pivot is essentially a collective all-reduce operation involving all the processes in this process column, as the processes must collectively determine the pivot row and receive a copy of this row. At the end of this phase, the column of panels will have been pivoted and $LU$ factored, yielding an $N\times \NB$ lower triangular matrix $L$ distributed between the column of processes. 

Once panel factorization is completed, each of the processes which participated in the factorization packs their section of the $L$ matrix, along with some index data conveying the pivoting information, into a buffer and broadcasts this buffer to all other processes in their row of the process grid. This step is graphically shown in Figure \ref{fig:lbcast}. No computation is performed in this step, and only the data in the buffer holding $L$ is accessed or modified. As the $L$ matrix is typically large for the majority of the HPL benchmark, the performance of this phase is heavily dependent on the amount of bandwidth available for inter-process communication, as well as the efficiency of the broadcast algorithm used.

With the $L$ matrix and pivoting information broadcast to all processes, the final major communication phase in each process is to apply all the row pivots determined in FACT to the remainder of the $A$ matrix on each process, and collectively construct the $U$ matrix which resides to the right of the $\NB \times \NB$ currently factored panel. Since the full set of $\NB$ pivots are known in this phase, we can perform the required communication in bulk via routines equivalent to \verb|MPI_Scatterv| followed by \verb |MPI_Allgatherv|. Each process first assembles their rows to be communicated into buffers, which requires an irregular access of their local $A$ matrix. Each process in the process row containing the currently factored panel then scatter the $\NB$ source rows to their destination processes in each process column via a \verb|Scatterv| communication. Following this, all processes in a column collectively assemble their section of the distributed $\NB \times N$ matrix, $U$, via an \verb|Allgatherv| communication. The data accessed and the communication directions are shown graphically in Figure \ref{fig:laswp}.

The final phase of the iteration is the most computationally demanding but requires no inter-process communication. The pivoted rows of the global matrix have been assembled into the $U$ matrix, and the computations from FACT are extended to these rows and applied as a single DTRSM routine, using the low-triangular piece of the factored diagonal panel. With the $L$ and $U$ matrices constructed and duplicated along the process rows and columns, respectively, the last computation is to apply a rank $\NB$ update to the trailing sub-matrix of $A$ distributed among all the processes. This computation is a distributed $N\times N \times \NB$ DGEMM which subtracts the product $LU$ from the trailing sub-matrix of $A$. The data accessed for this computation is shown graphically in Figure \ref{fig:update}.

\section{rocHPL Design}

AMD's implementation of the HPL benchmark, named \verb|rocHPL| \cite{rochpl}, is based on the open-source HPL implementation hosted on Netlib \cite{netlib-hpl}. This reference HPL code is parallelized with MPI, but otherwise contains no other parallel programming model. Our modifications to this HPL implementation involved adding GPU support via AMD's ROCm platform, runtime, and toolchains. The \verb|rocHPL| code is written using the HIP programming language and leverages linear algebra routines highly optimized for AMD's latest discrete GPUs via the rocBLAS math library.

As noted above, the recent works of Tan et al. \cite{tan2021optimizing} and Kim et al. \cite{kim2022snuhpl} both argue that the computational throughput of modern accelerators is so large that the entire matrix $A$ must be stored in the accelerators' high-bandwidth memory (HBM), as moving data from/to CPU memory would be too costly. As the AMD Instinct MI250X accelerators contain specialized hardware accelerating the crucial DGEMM computations, computational throughput has been even further increased beyond even what these works consider. We must therefore follow a similar design in \verb|rocHPL|, storing the matrix $A$ across each of the MI250X's 128 GB HBM capacity. 

It then becomes natural to consider whether all phases of the HPL benchmark should be performed on the accelerator, with the host process only serving to coordinate MPI communication. The UPDATE phase is, of course, a natural fit for the accelerator's high computational throughput. Likewise, the LBCAST and RS phases map easily to the accelerators as the required local data motion for row-swapping is accelerated using the GPUs' high memory bandwidth, and MPI communications can leverage both the high-bandwidth Infinity Fabric links between GPUs as well as the direct connection of the network interface cards (NICs) to the GPUs on node. The FACT phase, however, remains a challenge to execute on the accelerator. While it is true that many of the individual BLAS computations in FACT would be accelerated on the GPU, the communications required for row-pivoting would require frequent host-device synchronizations and would consequently introduce significant amounts of GPU idle time due to kernel launch latency. We therefore follow a similar approach to that of Tan et al. and Kim et al. and transfer necessary data back to the host processes in order for the FACT computations to be performed on the CPU before sending needed data back to the accelerators.

Fortunately, performing the FACT phase on the CPU leads to a relatively simple way to hide some necessary MPI communication by local computation using the `look-ahead' mechanism in the HPL benchmark. By noting that the FACT phase of each iteration requires only the next $\NB$ columns of the matrix, the look-ahead works by splitting the UPDATE phase on each process which will be performing the FACT phase in the next iteration. These processes first perform the UPDATE phase on only the leading $\NB$ columns, and then immediately begin transferring these columns to the host for factorization while completing the UPDATE phase on the remaining local matrix. This approach leads to an iteration whose timeline of execution looks similar to that shown in Figure \ref{fig:timeline}. When the UPDATE phase begins, the computations are performed on just the look-ahead first so that this section of columns can be transferred to the CPU, and then transferred back after the FACT phase. Once the FACT data arrives back on the GPU, the LBCAST communication can be done all while the remaining trailing UPDATE is being completed on the GPU. Processes which do not participate in the FACT phase simply wait in the LBCAST phase. After the UPDATE phase is completed, the row pivots computed in FACT are applied which requires a GPU kernel to gather the rows to be communicated, followed by MPI communication, and a GPU kernel to scatter the received rows back into $A$.

\subsection{Multi-threaded Panel Factorization}
At the beginning of the HPL benchmark, the computational work on the accelerator in each iteration can effectively hide both transfers to and from the host for the FACT computation, as well as the LBCAST communication. But as the benchmark progresses, the amount of work being performed in the UPDATE phase decreases until it is no longer able to hide these other phases. In order to maximize the duration of the benchmark where communication and factorization are hidden by UPDATE, and to spend the minimal amount of time without the UPDATE phase on the critical path, it is crucial to perform the FACT phase as fast as possible on the CPU.

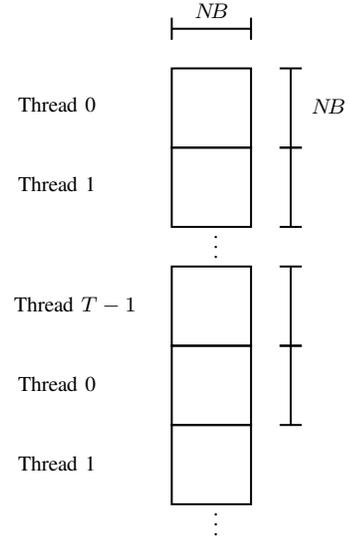
\begin{figure}[tbp]
\centering
\tikzset{every picture/.style={line width=0.75pt}} 

\begin{tikzpicture}[x=0.75pt,y=0.75pt,yscale=-1,xscale=1]

\draw   (120,40) -- (160,40) -- (160,80) -- (120,80) -- cycle ;
\draw   (120,80) -- (160,80) -- (160,120) -- (120,120) -- cycle ;
\draw   (120,140) -- (160,140) -- (160,180) -- (120,180) -- cycle ;
\draw   (120,180) -- (160,180) -- (160,220) -- (120,220) -- cycle ;
\draw   (120,220) -- (160,220) -- (160,260) -- (120,260) -- cycle ;
\draw    (120,20) -- (160,20) ;
\draw [shift={(160,20)}, rotate = 180] [color={rgb, 255:red, 0; green, 0; blue, 0 }  ][line width=0.75]    (0,5.59) -- (0,-5.59)   ;
\draw [shift={(120,20)}, rotate = 180] [color={rgb, 255:red, 0; green, 0; blue, 0 }  ][line width=0.75]    (0,5.59) -- (0,-5.59)   ;
\draw    (180,40) -- (180,80) ;
\draw [shift={(180,80)}, rotate = 270] [color={rgb, 255:red, 0; green, 0; blue, 0 }  ][line width=0.75]    (0,5.59) -- (0,-5.59)   ;
\draw [shift={(180,40)}, rotate = 270] [color={rgb, 255:red, 0; green, 0; blue, 0 }  ][line width=0.75]    (0,5.59) -- (0,-5.59)   ;
\draw    (180,80) -- (180,120) ;
\draw [shift={(180,120)}, rotate = 270] [color={rgb, 255:red, 0; green, 0; blue, 0 }  ][line width=0.75]    (0,5.59) -- (0,-5.59)   ;
\draw [shift={(180,80)}, rotate = 270] [color={rgb, 255:red, 0; green, 0; blue, 0 }  ][line width=0.75]    (0,5.59) -- (0,-5.59)   ;
\draw    (180,140) -- (180,180) ;
\draw [shift={(180,180)}, rotate = 270] [color={rgb, 255:red, 0; green, 0; blue, 0 }  ][line width=0.75]    (0,5.59) -- (0,-5.59)   ;
\draw [shift={(180,140)}, rotate = 270] [color={rgb, 255:red, 0; green, 0; blue, 0 }  ][line width=0.75]    (0,5.59) -- (0,-5.59)   ;
\draw    (180,180) -- (180,220) ;
\draw [shift={(180,220)}, rotate = 270] [color={rgb, 255:red, 0; green, 0; blue, 0 }  ][line width=0.75]    (0,5.59) -- (0,-5.59)   ;
\draw [shift={(180,180)}, rotate = 270] [color={rgb, 255:red, 0; green, 0; blue, 0 }  ][line width=0.75]    (0,5.59) -- (0,-5.59)   ;

\draw (146,122) node [anchor=north west][inner sep=0.75pt]  [font=\footnotesize,rotate=-90] [align=left] {$\displaystyle \cdots $};
\draw (146,262) node [anchor=north west][inner sep=0.75pt]  [font=\footnotesize,rotate=-90] [align=left] {$\displaystyle \cdots $};
\draw (130,6) node [anchor=north west][inner sep=0.75pt]  [font=\footnotesize] [align=left] {$\displaystyle N\!B$};
\draw (189,54) node [anchor=north west][inner sep=0.75pt]  [font=\footnotesize] [align=left] {$\displaystyle N\!B$};
\draw (41,53) node [anchor=north west][inner sep=0.75pt]  [font=\footnotesize] [align=left] {Thread 0};
\draw (41,93) node [anchor=north west][inner sep=0.75pt]  [font=\footnotesize] [align=left] {Thread 1};
\draw (39,154) node [anchor=north west][inner sep=0.75pt]  [font=\footnotesize] [align=left] {Thread $\displaystyle T-1$};
\draw (41,194) node [anchor=north west][inner sep=0.75pt]  [font=\footnotesize] [align=left] {Thread 0};
\draw (41,234) node [anchor=north west][inner sep=0.75pt]  [font=\footnotesize] [align=left] {Thread 1};

\end{tikzpicture}
\caption{Graphical depiction of the multi-threading strategy used in the FACT computation on CPU. The ensemble of $\NB$ columns is split into $\NB \times \NB$ tiles, and the tiles are round-robined between $T$ CPU threads.}
\label{fig:pfact-threaded}
\end{figure}

The design in \verb|rocHPL| is to let every MPI process manage one and only one GPU device. In the case of MI250X GPUs, where each GCD of the module presents to the OS as a distinct GPU, each MPI rank manages a unique GCD. Assuming each MPI rank is also bound to a distinct CPU core, this leaves potentially many unused CPU cores which can be leveraged in the FACT phase through multi-threading. While many CPU BLAS libraries offer multi-threaded implementations of computationally expensive BLAS routines, such as the DGEMMs and DTRSMs needed in FACT, we opt instead to multi-thread the entirety of the FACT phase by manually distributing the computation among CPU threads. 

The matrix being LU factored in the FACT phase is tall and skinny. It consists of only $\NB$ columns, but potentially many thousands of rows. This makes it amenable to parallelization by distributing chunks of rows between threads on the host, an approach similar to the technique of Parallel Cache Assignment (PCA) by Castaldo et al. \cite{castaldo2010scaling} and the work on parallel panel factorization by Dongarra et al. \cite{dongarra2014achieving} and Kurzak et al. \cite{kurzak2019linear}. At the beginning of the FACT phase, we create an OpenMP parallel region of $T$ threads and distribute work between threads by blocking the tall and skinny matrix into tiles of $\NB$ rows, assigning blocks in a round-robin fashion to each thread, as shown graphically in Figure \ref{fig:pfact-threaded}. We choose square tile sizes purely out of convenience as this way the first tile, which will contain the upper-triangular factor as well as all the source rows during pivoting, is guaranteed to be assigned to the main thread. 

The original Netlib HPL code on which \verb|rocHPL| is based contains several serial and blocked LU factorization methods, including left-looking, right-looking and Crout factorizations. Each of these is directly parallelizable with the tiling strategy. The determination of the pivot row is implemented as a parallel reduction over all OpenMP threads, after which the main thread calls MPI to complete the reduction across all processes in the process column. The main thread then applies the row pivot and synchronizes with the remaining threads so that the rank-1 update to the trailing sub-matrix can be applied in a parallel fashion using all the threads. For the blocked factorization methods, a similar idea is applied where the main thread performs the DTRSM updates to the upper-triangular factor, after which each thread uses the result to perform their section of the trailing update. With this approach, the data in each tile is accessed by only one thread, with the exception of any accesses by the main thread when applying row pivots. The data can therefore be kept resident in the CPU caches near the physical core to which that thread is bound. In addition, using the 64-core AMD CPUs on Frontier the entirety of the data accessed during the FACT phase typically remains resident in the L3 cache.

\begin{figure}[tbp]
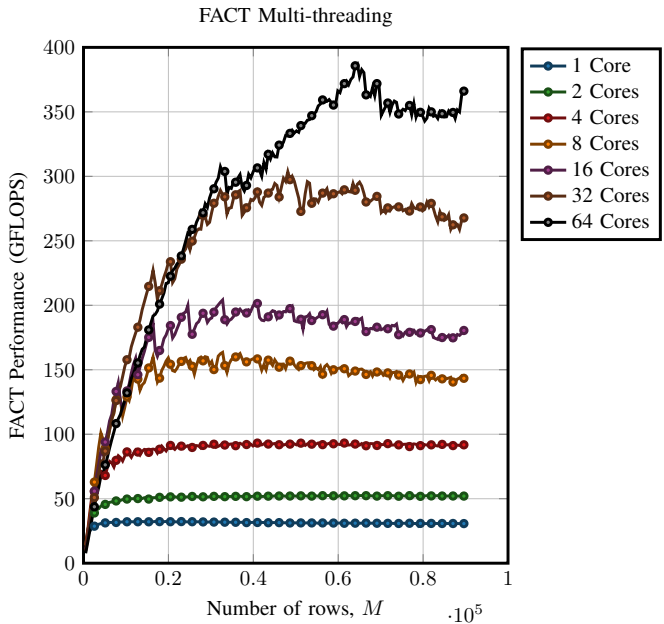

\centering
\input Figures/pfact-mptest
\caption{Multi-threading performance test for FACT phase. Performance in GFLOPS is measured when factoring an $M\times \NB$ matrix on a single process for $\NB = 512$ and $M$ a range of multiples of $\NB$. Different curves show different numbers of CPU cores used in powers of 2 from 1 to 64.}
\label{fig:pfact-mptest}
\end{figure}

To demonstrate the performance benefits of this multi-threading strategy in the FACT phase, we show the results of a performance test in Figure \ref{fig:pfact-mptest} on a single Frontier node using BLIS v4.0 as the CPU BLAS library. The performance of the FACT phase when factoring an $M \times \NB$ matrix is measured for $\NB = 512$ and $M$ taken to be various multiples of $\NB$. We run this test with a single process to eliminate time which would be spent by the main thread determining and exchanging pivots with MPI. The factorization algorithm used is the recursive right-looking with two subdivisions in the recursion and a base block size of 16. On the base block, the factorization algorithm used is a right-looking factorization. We execute the FACT computation across these problem sizes using different numbers of CPU cores. From the figure, we see that the performance of the FACT phase is considerably improved through multi-threading and that using large numbers of CPU cores benefits performance for even the relatively small problem sizes. 

\subsection{CPU Core Time Sharing}
With the multi-threading strategy for the FACT phase, an important consideration is where to place CPU threads to maximize the performance in each FACT computation. Consider the example of the Frontier node architecture; as there are eight GCDs present in the node, we launch eight MPI processes and bind each process to the CCD that is nearest to the GCD it will manage (c.f. the node topology diagram in \cite{crusherquickstart}). A natural choice then is to have each process create seven additional OpenMP threads when it enters the FACT phase, so that the process can leverage all eight CPU cores in its CCD. 

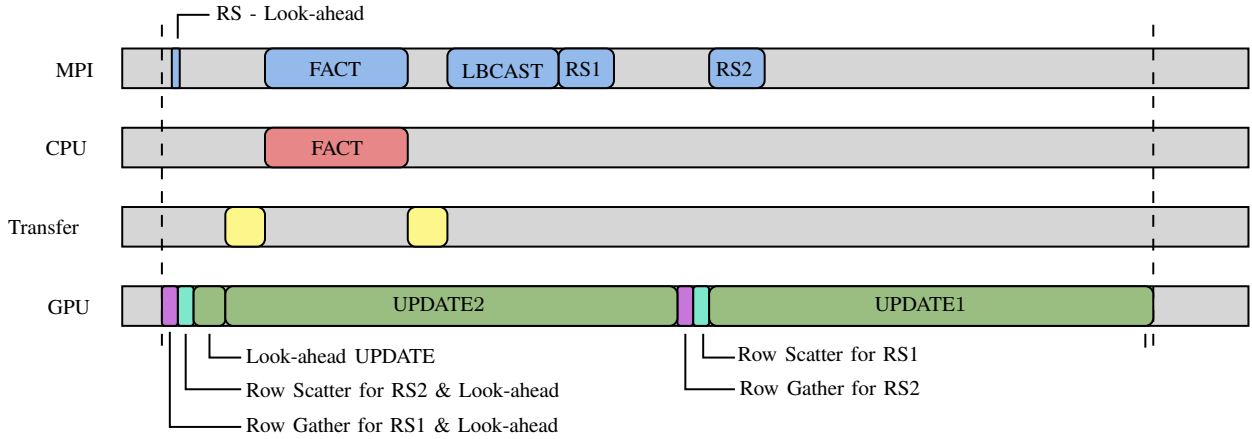
\begin{figure*}[tbp]
\centering
\tikzset{every picture/.style={line width=0.75pt}} 

\begin{tikzpicture}[x=0.75pt,y=0.75pt,yscale=-1,xscale=1]

\draw  [fill={rgb, 255:red, 214; green, 214; blue, 214 }  ,fill opacity=1 ] (92,81) -- (660,81) -- (660,101) -- (92,101) -- cycle ;
\draw  [fill={rgb, 255:red, 214; green, 214; blue, 214 }  ,fill opacity=1 ] (92,121) -- (660,121) -- (660,141) -- (92,141) -- cycle ;
\draw  [fill={rgb, 255:red, 214; green, 214; blue, 214 }  ,fill opacity=1 ] (92,161) -- (660,161) -- (660,181) -- (92,181) -- cycle ;
\draw  [fill={rgb, 255:red, 214; green, 214; blue, 214 }  ,fill opacity=1 ] (92,41) -- (660,41) -- (660,61) -- (92,61) -- cycle ;
\draw  [fill={rgb, 255:red, 231; green, 135; blue, 135 }  ,fill opacity=1 ] (164,85) .. controls (164,82.79) and (165.79,81) .. (168,81) -- (232,81) .. controls (234.21,81) and (236,82.79) .. (236,85) -- (236,97) .. controls (236,99.21) and (234.21,101) .. (232,101) -- (168,101) .. controls (165.79,101) and (164,99.21) .. (164,97) -- cycle ;
\draw  [fill={rgb, 255:red, 201; green, 118; blue, 218 }  ,fill opacity=1 ] (372,162.6) .. controls (372,161.72) and (372.72,161) .. (373.6,161) -- (378.4,161) .. controls (379.28,161) and (380,161.72) .. (380,162.6) -- (380,179.4) .. controls (380,180.28) and (379.28,181) .. (378.4,181) -- (373.6,181) .. controls (372.72,181) and (372,180.28) .. (372,179.4) -- cycle ;
\draw  [fill={rgb, 255:red, 126; green, 231; blue, 204 }  ,fill opacity=1 ] (120,162.6) .. controls (120,161.72) and (120.72,161) .. (121.6,161) -- (126.4,161) .. controls (127.28,161) and (128,161.72) .. (128,162.6) -- (128,179.4) .. controls (128,180.28) and (127.28,181) .. (126.4,181) -- (121.6,181) .. controls (120.72,181) and (120,180.28) .. (120,179.4) -- cycle ;
\draw  [fill={rgb, 255:red, 154; green, 190; blue, 129 }  ,fill opacity=1 ] (128,164.2) .. controls (128,162.43) and (129.43,161) .. (131.2,161) -- (140.8,161) .. controls (142.57,161) and (144,162.43) .. (144,164.2) -- (144,177.8) .. controls (144,179.57) and (142.57,181) .. (140.8,181) -- (131.2,181) .. controls (129.43,181) and (128,179.57) .. (128,177.8) -- cycle ;
\draw  [fill={rgb, 255:red, 154; green, 190; blue, 129 }  ,fill opacity=1 ] (144,165) .. controls (144,162.79) and (145.79,161) .. (148,161) -- (368,161) .. controls (370.21,161) and (372,162.79) .. (372,165) -- (372,177) .. controls (372,179.21) and (370.21,181) .. (368,181) -- (148,181) .. controls (145.79,181) and (144,179.21) .. (144,177) -- cycle ;
\draw  [fill={rgb, 255:red, 148; green, 185; blue, 235 }  ,fill opacity=1 ] (312,45) .. controls (312,42.79) and (313.79,41) .. (316,41) -- (336,41) .. controls (338.21,41) and (340,42.79) .. (340,45) -- (340,57) .. controls (340,59.21) and (338.21,61) .. (336,61) -- (316,61) .. controls (313.79,61) and (312,59.21) .. (312,57) -- cycle ;
\draw  [fill={rgb, 255:red, 253; green, 246; blue, 139 }  ,fill opacity=1 ] (144,125) .. controls (144,122.79) and (145.79,121) .. (148,121) -- (160,121) .. controls (162.21,121) and (164,122.79) .. (164,125) -- (164,137) .. controls (164,139.21) and (162.21,141) .. (160,141) -- (148,141) .. controls (145.79,141) and (144,139.21) .. (144,137) -- cycle ;
\draw  [fill={rgb, 255:red, 253; green, 246; blue, 139 }  ,fill opacity=1 ] (236,125) .. controls (236,122.79) and (237.79,121) .. (240,121) -- (252,121) .. controls (254.21,121) and (256,122.79) .. (256,125) -- (256,137) .. controls (256,139.21) and (254.21,141) .. (252,141) -- (240,141) .. controls (237.79,141) and (236,139.21) .. (236,137) -- cycle ;
\draw  [fill={rgb, 255:red, 148; green, 185; blue, 235 }  ,fill opacity=1 ] (164,45) .. controls (164,42.79) and (165.79,41) .. (168,41) -- (232,41) .. controls (234.21,41) and (236,42.79) .. (236,45) -- (236,57) .. controls (236,59.21) and (234.21,61) .. (232,61) -- (168,61) .. controls (165.79,61) and (164,59.21) .. (164,57) -- cycle ;
\draw  [fill={rgb, 255:red, 148; green, 185; blue, 235 }  ,fill opacity=1 ] (256,45) .. controls (256,42.79) and (257.79,41) .. (260,41) -- (308,41) .. controls (310.21,41) and (312,42.79) .. (312,45) -- (312,57) .. controls (312,59.21) and (310.21,61) .. (308,61) -- (260,61) .. controls (257.79,61) and (256,59.21) .. (256,57) -- cycle ;
\draw  [dash pattern={on 4.5pt off 4.5pt}]  (112,29) -- (112,193) ;
\draw  [dash pattern={on 4.5pt off 4.5pt}]  (612,29) -- (612,193) ;
\draw   (152,197) -- (136,197) -- (136,185) ;
\draw   (152,213) -- (124,213) -- (124,185) ;
\draw  [fill={rgb, 255:red, 126; green, 231; blue, 204 }  ,fill opacity=1 ] (380,162.6) .. controls (380,161.72) and (380.72,161) .. (381.6,161) -- (386.4,161) .. controls (387.28,161) and (388,161.72) .. (388,162.6) -- (388,179.4) .. controls (388,180.28) and (387.28,181) .. (386.4,181) -- (381.6,181) .. controls (380.72,181) and (380,180.28) .. (380,179.4) -- cycle ;
\draw  [fill={rgb, 255:red, 154; green, 190; blue, 129 }  ,fill opacity=1 ] (388,165) .. controls (388,162.79) and (389.79,161) .. (392,161) -- (608,161) .. controls (610.21,161) and (612,162.79) .. (612,165) -- (612,177) .. controls (612,179.21) and (610.21,181) .. (608,181) -- (392,181) .. controls (389.79,181) and (388,179.21) .. (388,177) -- cycle ;
\draw  [fill={rgb, 255:red, 201; green, 118; blue, 218 }  ,fill opacity=1 ] (112,162.6) .. controls (112,161.72) and (112.72,161) .. (113.6,161) -- (118.4,161) .. controls (119.28,161) and (120,161.72) .. (120,162.6) -- (120,179.4) .. controls (120,180.28) and (119.28,181) .. (118.4,181) -- (113.6,181) .. controls (112.72,181) and (112,180.28) .. (112,179.4) -- cycle ;
\draw  [fill={rgb, 255:red, 148; green, 185; blue, 235 }  ,fill opacity=1 ] (388,45) .. controls (388,42.79) and (389.79,41) .. (392,41) -- (412,41) .. controls (414.21,41) and (416,42.79) .. (416,45) -- (416,57) .. controls (416,59.21) and (414.21,61) .. (412,61) -- (392,61) .. controls (389.79,61) and (388,59.21) .. (388,57) -- cycle ;
\draw   (401,197) -- (385,197) -- (385,185) ;
\draw   (401,213) -- (376,213) -- (376,185) ;
\draw    (608,185) -- (608,192) ;
\draw  [fill={rgb, 255:red, 148; green, 185; blue, 235 }  ,fill opacity=1 ] (117,41.8) .. controls (117,41.36) and (117.36,41) .. (117.8,41) -- (120.2,41) .. controls (120.64,41) and (121,41.36) .. (121,41.8) -- (121,60.2) .. controls (121,60.64) and (120.64,61) .. (120.2,61) -- (117.8,61) .. controls (117.36,61) and (117,60.64) .. (117,60.2) -- cycle ;
\draw   (136,24) -- (120,24) -- (120,36) ;
\draw   (152,232) -- (116,232) -- (116,184) ;

\draw (185,46) node [anchor=north west][inner sep=0.75pt]  [font=\footnotesize] [align=left] {FACT};
\draw (262,47) node [anchor=north west][inner sep=0.75pt]  [font=\footnotesize] [align=left] {LBCAST};
\draw (314,46) node [anchor=north west][inner sep=0.75pt]  [font=\footnotesize] [align=left] {RS1};
\draw (186,86) node [anchor=north west][inner sep=0.75pt]  [font=\footnotesize] [align=left] {FACT};
\draw (227,165) node [anchor=north west][inner sep=0.75pt]  [font=\footnotesize] [align=left] {UPDATE$\displaystyle 2$};
\draw (57,47) node [anchor=north west][inner sep=0.75pt]  [font=\footnotesize] [align=left] {MPI};
\draw (52,86) node [anchor=north west][inner sep=0.75pt]  [font=\footnotesize] [align=left] {CPU};
\draw (33,126) node [anchor=north west][inner sep=0.75pt]  [font=\footnotesize] [align=left] {Transfer};
\draw (53,166) node [anchor=north west][inner sep=0.75pt]  [font=\footnotesize] [align=left] {GPU};
\draw (153,191) node [anchor=north west][inner sep=0.75pt]  [font=\footnotesize] [align=left] {Look-ahead UPDATE};
\draw (153,208) node [anchor=north west][inner sep=0.75pt]  [font=\footnotesize] [align=left] {Row Scatter for RS2 \& Look-ahead};
\draw (470,165) node [anchor=north west][inner sep=0.75pt]  [font=\footnotesize] [align=left] {UPDATE$\displaystyle 1$};
\draw (390,46) node [anchor=north west][inner sep=0.75pt]  [font=\footnotesize] [align=left] {RS2};
\draw (401,190) node [anchor=north west][inner sep=0.75pt]  [font=\footnotesize] [align=left] {Row Scatter for RS1};
\draw (402,207) node [anchor=north west][inner sep=0.75pt]  [font=\footnotesize] [align=left] {Row Gather for RS2};
\draw (153,226) node [anchor=north west][inner sep=0.75pt]  [font=\footnotesize] [align=left] {Row Gather for RS1 \& Look-ahead};
\draw (138,18) node [anchor=north west][inner sep=0.75pt]  [font=\footnotesize] [align=left] {RS - Look-ahead};

\end{tikzpicture}
\caption{Diagram of the execution of a single iteration in HPL's factorization with the split update formulation. When the execution time of the UPDATE phase is large, we observe that computation on the GPU can effectively hide all other phases in the HPL iteration.}
\label{fig:timeline-split}
\end{figure*}

However, it is often possible for a process to utilize even more CPU cores than would be available through a simple partitioning of all available cores. Consider a 2D process grid of $P\times Q = 2 \times 4$ on the Frontier node architecture. At any given iteration of the computation in HPL only two processes will coordinate on computing the panel factorization while the other six processes are waiting to receive the LBCAST. If both processes performing the panel factorization each use eight CPU cores while the remaining six MPI processes each use a single CPU core, that leaves 42 idle CPU cores on the socket during this iteration. As the benchmark proceeds through iterations, the 16 total CPU cores being used during each FACT phase will cycle between different CCDs, but we will still always observe 42 total idle CPU cores in every iteration. With this observation, we consider a generic way to leverage all CPU cores in each HPL iteration by over-subscribing OpenMP threads to physical CPU cores.

In the general case of launching a node-local $P\times Q$ process grid to a node with $C$ CPU cores, we bind each of the processes to a distinct root core and consider the remaining $\bar{C} = C - PQ$ cores as a pool of resources. This pool is partitioned into $P$ non-overlapping groups, each with $\frac{\bar{C}}{P}$ cores, and each group is assigned to a distinct process row. Every MPI rank in each process column then uses OpenMP bindings to specify a total of $T = 1+ \frac{\bar{C}}{P}$ OpenMP threads and binds them to its root core and its process row's partition of the pool. In this way, every FACT phase will leverage a total of $PT = P + \bar{C}$ CPU cores on the node. In the extreme case of a $P\times 1$ local process grid on the node, this core binding reduces to a simple partitioning of available CPU cores, as all processes on the node must participate in the FACT phase simultaneously. At the opposite extreme of a $1 \times Q$ local process grid on the node, the amount of CPU core sharing is maximized since at most one process on the node will ever be computing the FACT phase at any given time. 

In \verb|rocHPL| we have implemented a generic wrapper script to compute these OpenMP bindings when launching the benchmark. The CPU core time sharing, as well as the decomposition of the global problem among the compute nodes, uses input from the user which describes the local process grid configuration desired on each node.

\subsection{Split Update}
Examination of the timeline view of execution in Figure \ref{fig:timeline} shows that the division of work between the host CPU and accelerator allows us to effectively hide the FACT and LBCAST phases behind the local UPDATE computation. However, the communication time required to perform the RS phase still leads to idle time on the accelerator. It is therefore advantageous to consider ways to hide this communication time with local computation as well. 

A simple way to accomplish this communication hiding would be to divide the RS and UPDATE phases column-wise into several smaller chunks and apply a pipe-lining strategy. In this way, the local computation to perform the UPDATE phase on a chunk can hide the communication time for the RS phase for the next chunk. This strategy is what is applied in Tan et al. \cite{tan2021optimizing} and Kim et al. \cite{kim2022snuhpl}, who both use a multi-threaded implementation to coordinate the different chunks and different phases. Such a multi-threaded strategy is costly in our HPL implementation, however, as using multiple threads to pipe-line different phases utilizes CPU cores that could otherwise be used in FACT.

We opt instead for an alternative way to hide the communication time in the RS phase that requires no addition multi-threading, which we call a `split update'. Let us denote by $n$ the number of columns in the local section of $A$ on a process at the start of the HPL benchmark. Note that due to the 2D block-cyclic distribution of $A$, $n$ will be the same for all processes in each process column. Consider splitting the local matrix column-wise into two pieces with $n_1$ and $n_2$ columns which we call the `left' and `right' sections of the local $A$ matrix, respectively. We select $n_1$ such that it is a multiple of $\NB$. We denote by UPDATE1 and UPDATE2, the application of the UPDATE phase on the left and right sections, respectively, and likewise for the RS phase. The idea of the split update is to use the UPDATE computation on one section to hide the MPI communication of the RS phase of the other section. The key observation is that in order to do so, the needed rows for the RS stage on one section must be gathered before the UPDATE of the other section is started.

The split update formulation leads to a timeline of execution that resembles the one shown in Figure \ref{fig:timeline-split}. At the start of an iteration, we assume that RS2 communication has already been completed. That is, we assume that the rows in the right section of the local $A$ matrix have been communicated, though not necessarily scattered back into $A$. We begin the iteration by gathering the needed rows for communication from both the look-ahead and left section and scattering the communicated rows from the right section back into $A$. While the rows are scattered, the communication of rows for only the look-ahead is performed, and the received rows are written into the look-ahead. The iteration then proceeds as it does without the split update, with the UPDATE phase being performed on the look-ahead and the result copied back to the host for the FACT phase and subsequent LBCAST. While the transfers, FACT, and LBCAST are executed, the UPDATE2 phase is computed on the accelerator. However, since the rows of the left section of $A$ have already been gathered for communication, the RS1 communication can also be performed at this point and be hidden by UPDATE2. Following UPDATE2, the rows for the next iteration in the right section of $A$ are gathered to prepare for the RS2 communication. The UPDATE1 phase can then be queued, first scattering the communicated rows back into $A$, and the RS2 communication can be hidden by this local computation. Note that if this process performed the FACT phase, then the number of columns updated in UPDATE1 will be $n_1 - \NB$.

Because of the staggered fashion in which the left and right updates are performed, interleaved with their respective row gathering and scattering, we must keep the number of columns in the right section of the local $A$ matrix, $n_2$, on each process fixed for each iteration while $n_1$ decreases. Since we pick $n_1$ to be a multiple of $\NB$, eventually $n_1$ will equal $\NB$, and the look-ahead will then eventually become the entirety of the remaining left section. After this occurs, there is no longer a split update formulation, and the iterations fall back to the form shown in Figure \ref{fig:timeline} where the RS communication is not hidden by UPDATE. 

For the split update formulation to effectively hide all communication time, the right section of the local $A$ matrix must be at least large enough to hide the data transfers to and from the host, as well as the FACT, LBCAST, and RS1 phases. It is then natural to ask: if the UPDATE2 phase can initially hide all time spent in these phases, will it continue to hide them as the benchmark progresses? To determine this, note that since $n_2$ remains fixed while $n_1$ decreases, the UPDATE2 phase is always updating the same number of columns of $A$ in each iteration for as long as $n_1$ remains non-zero. In terms of computational cost, while $n_2$ remains fixed the UPDATE2 computation scales linearly with the number of rows, denoted by $m$, in the local piece of $A$ being updated on this process. Likewise, the other phases hidden by UPDATE2, except for the RS1 phase, share this linear scaling with $m$. Indeed, the data transfers to/from the host, the FACT phase, and the LBCAST phase each have a linear complexity in the number of local rows of $A$ being updated in each iteration. The row-swapping communication in RS1, on the other hand, has a complexity which is linear in $n_1$, and is therefore decreasing as $n_1$ decreases. That rate is roughly the same rate as the local number of rows of $A$ decreases. We can therefore conclude that if the UPDATE2 can initially hide each of these components, it will continue to hide them up until the point when the left section has decreased to zero columns and the RS2 communication can no longer be hidden. In practice, we have observed that this split update formulation is able to hide all MPI communication by UPDATE phases for approximately 75\% of the execution time of the HPL benchmark on a single Frontier node. 

Since it is crucial that $n_2$ be chosen only large enough to hide the FACT, LBCAST, and RS1 phases, its selection at the beginning of the benchmark is a key consideration in the split update formulation. While some performance models could be used to estimate the optimal size of $n_2$, we instead allow the user to input a `split fraction' parameter to \verb|rocHPL| to indicate what percentage of columns should be in the right section of $A$ and leave this input as a tuning parameter. For HPL runs on a single Frontier node, we typically find that splitting the local $A$ matrix in half between the left and right sections works optimally.

\section{Performance Results}
In this section we present some performance results of the \verb|rocHPL| benchmark on a single node, and scaled to multiple nodes, of the Crusher system at the Oak Ridge Leadership Compute Facility (OLCF). Crusher is a Frontier early access cluster and therefore shares the same node architecture as Frontier. Crusher is an HPE Cray EX supercomputer system with each node consisting of a single socket optimized 3rd Gen EPYC 64 core processor, four AMD Instinct MI250X accelerators, and four HPE Slingshot 200Gbps network interfaces, each directly attached to a distinct MI250X GPU. For all performance results below, we use GCC v11.2.0 as our C++ compiler, ROCm v5.4.0 to compile HIP kernels, and use the rocBLAS v2.46 GPU BLAS library distributed with ROCm v5.4.0. For the CPU BLAS library, we use BLIS v4.0. Finally, we use Cray-MPICH v8.1.17 as the MPI implementation. 

\begin{figure}[tbp]
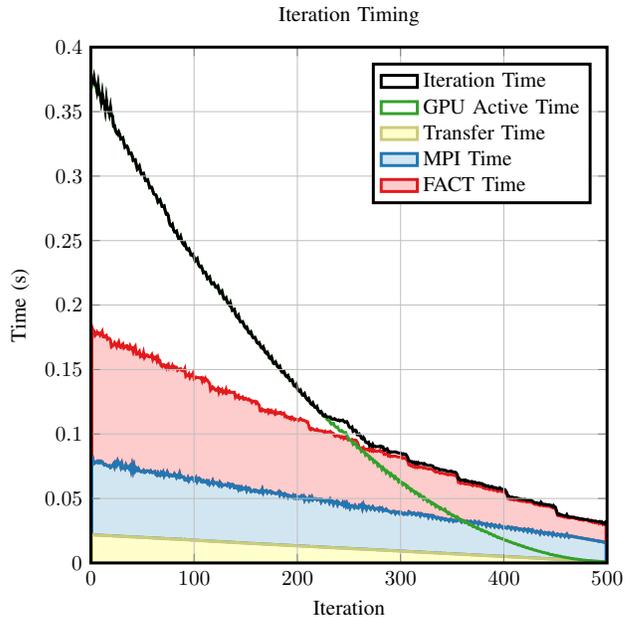

\centering
\input Figures/iteration-time
\caption{Timing breakdown of each iteration in a run of the HPL benchmark on a single node of Crusher. The black line shows the total time at each iteration and the green line shows the time the GPU was active during the iteration. The stacked areas of red, blue, and yellow show the FACT computation time, the MPI communication time, and the host-device data transfer time at each iteration, respectively. }
\label{fig:iteration}
\end{figure}

\subsection{Single Node Performance}
Beginning with a single node run on Crusher, we execute the \verb|rocHPL| benchmark by launching eight processes in a $P\times Q = 4\times 2$ process grid to a single node. We use a global problem size of $N=256,000$ which, combined with necessary workspace buffers, effectively fills the HBM capacity of each of the four MI250X GPUs on the node. As is typical of HPL implementations, the choice of blocking factor $\NB$ is an important balance of computation and communication performance. The block size $\NB$ should be chosen at least large enough that the large DGEMM computations reach a high percentage of peak performance on the device, while choosing $\NB$ to be as small as possible allows for maximal overlap of communication and computation in each iteration. For the Frontier node architecture, we typically choose $\NB = 512$ to strike this balance. At $\NB=512$ the DGEMMs required in HPL typically achieve 49 TFLOPS of performance on each MI250X GPU using the highly tuned DGEMM kernels available in rocBLAS. With this $\NB$ value, we also utilize a 50-50 left-right split in the split update formulation to hide MPI communication for row-swapping. With these parameters, the single node execution of the \verb|rocHPL| benchmark on Crusher achieves on average 153 TFLOPS of performance overall. 

We show in Figure \ref{fig:iteration} the timing breakdown of each iteration in the single node run of the \verb|rocHPL| benchmark. For each iteration, the process which owns the current diagonal panel records several timers. We record the overall iteration time along with several other components, namely: the total time the GPU spent actively computing in this iteration, the total time spent sending data to and from the host, the total time spent in MPI communication, and the total time spent computing the FACT phase on the CPU. We plot both the per-iteration time along with the total GPU active time in this figure. The remaining three timers are shown in the figure as stacked lines to show the critical path of execution near the end of the benchmark. 

From the behavior of the per-iteration time in Figure \ref{fig:iteration} we see two distinct regimes during the HPL benchmark execution. At the beginning of the benchmark, the per-iteration time precisely corresponds to the total GPU time in each iteration. This demonstrates that all other phases, including panel factorization and all MPI communication, are entirely hidden by GPU actively. Furthermore, 95\% of the GPU active time in each iteration is typically spent inside DGEMM computations. We therefore achieve a high percentage of the achievable computational throughput of the node in this regime. Indeed, as each large DGEMM computation achieves 49 TFLOPS of performance, we have an absolute limit of $4 \times 49 = 196$ TFLOPS of computational throughput in this regime. The \verb|rocHPL| benchmark prints a variety of performance metrics during execution, from which we typically see the running throughput in this regime achieve 90\% of this limit, or 175 TFLOPS. 

Around iteration 250, the left section in the split update is too small to adequately hide the RS2 communication and the per-iteration time can be seen to be slightly above the GPU active time. Soon after this, the left section of the split update becomes zero, and the right section shrinks until GPU activity is no longer on the critical path at all. From the stacked line plots of the host-device transfer time, MPI communication time, and FACT computation time in Figure \ref{fig:iteration} we see that these combined phases become the critical path of execution for the remainder of the benchmark execution. It is in this tail section that the running computational throughput of the benchmark decreases substantially to its final value, as the time per iteration in this regime is no longer compute-bound, but rather latency and communication bound. Nevertheless, using the optimization described above, the HPL implementation in \verb|rocHPL| still achieves an overall performance of 78\% of the achievable $\NB=512$ DGEMM computation rate of 49 TFLOPS per MI250X on a single Crusher node. 

\subsection{Multi-node Scaling}

When weak scaled to multiple nodes, we expect the per-iteration time breakdown of the HPL run to follow a similar trend to that shown in Figure \ref{fig:iteration}, albeit with more total iterations as the node count grows. However, while the computational work in each GPU and CPU socket remains the same as the problem is scaled, the MPI communication time is expected to grow compared to the single node results. This is due to two factors. First, the inter-node bandwidth uses the network interfaces which have less peak bandwidth than the Infinity Fabric links between the GCDs on a single node. This affects bandwidth-sensitive communication operations like those in the LBCAST and RS phases. Second, the latency cost of communications is expected to increase as the node count grows. This is important in latency-sensitive communications like the individual row pivots in the FACT phase, which are essentially MPI collectives across an entire process column. 

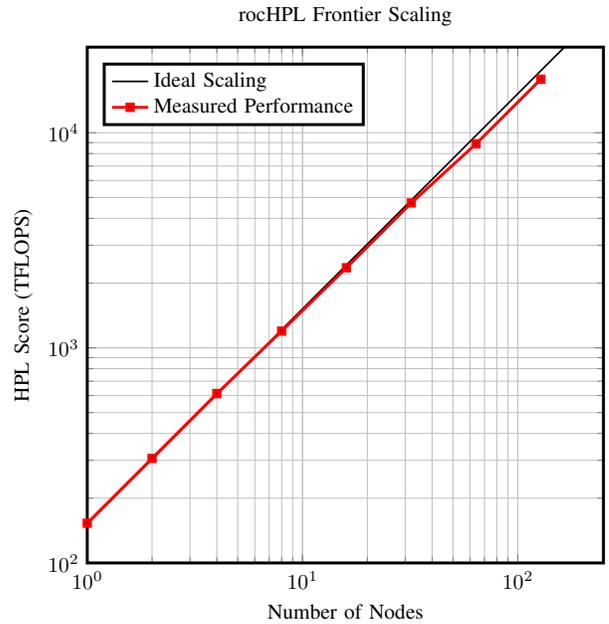
\begin{figure}[tbp]
\centering
\begin{tikzpicture}[scale=0.8]
\begin{axis}[
  xmode=log, 
  ymode=log, 
  grid=both, 
  major grid style={line width=.1pt,draw=gray!50}, 
  minor grid style={line width=.1pt,draw=gray!50}, 
  domain=1:8, 
  width=4in, 
  height=4in, 
  ymin=1e-16, 
  xlabel={Number of Nodes}, 
  ylabel={HPL Score (TFLOPS)}, 
  legend cell align=left, 
  legend pos=north west, 
  mark size=1.5pt, 
  line width=1.4pt, 
  legend entries={Ideal Scaling, Measured Performance},
  title={rocHPL Frontier Scaling},
  ymax=25000,
  ymin=100,
  xmin=1,
  xmax=250,
]

\addplot [domain=1:250, mark=none, thick ]
            { 151.8*x };
            
\addplot   table {
1  153.0
2  306.1
4  612.7
8  1196.3
16  2357.0
32  4716.4
64  8878.6
128  17746.0
};

\end{axis}
\end{tikzpicture}
\caption{Measured HPL score on multiple nodes of Crusher using the rocHPL benchmark. Benchmark is executed on $1, 2, 4, 8, \ldots, 128$ nodes.}
\label{fig:scaling}
\end{figure}

We show in Figure \ref{fig:scaling} the performance of the \verb|rocHPL| benchmark measured on $1, 2, 4, 8, \ldots, 128$ nodes of Crusher. We also show the ideal perfect weak scaling from the single node performance. For each node count, we keep the $P\times Q$ process grid square, or a grid with a 2:1 ratio of $P$ to $Q$. For the node-local process grid, which determines the amount of CPU core time sharing that we can perform, we maximize the number of process columns on-node. That is, once $Q$ is at least 8, we select the node-local process grid to be $1\times 8$. We scale the global problem size, $N$, to again fill the GPUs' HBM capacity, and hold $\NB$ fixed at 512, and the left-right split at 50\%, for all tests. We see in the figure that the \verb|rocHPL| benchmark scales very well to multiple nodes, achieving over 90\% weak-scaling efficiency from the single node score of 153 TFLOPS to the score of 17.75 PFLOPS on 128 nodes. Despite the relatively small node count for this test, this score would rank 38$^{\mathrm{th}}$ on the November 2022 Top500 list. 


\section{Discussion}
We have presented \verb|rocHPL|, AMD's open-source implementation of the HPL benchmark targeting accelerated node architectures designed for exascale systems. As with other recent works on leveraging modern accelerators in HPL, \verb|rocHPL| holds the entire problem in the accelerators' high-bandwidth memory and moves panels to the CPU only to perform the small latency-sensitive panel factorization. We detailed some performance optimizations used in \verb|rocHPL| including a multi-threading strategy for improving panel factorization performance on the CPU, a method for time-sharing CPU core resources between different processes on the same node, and a split update strategy that can effectively hide communication time required for performing row-pivoting. 

Detailed timing of the execution of \verb|rocHPL| on a single node of the Crusher system shows that our optimizations are able to entirely hide MPI communications and CPU work behind GPU compute activity for the first 50\% of the iterations in the benchmark. This, combined with the high performance DGEMM routines in rocBLAS, allows the benchmark to achieve a high percentage of the effective DGEMM computational throughput on each accelerator. Towards the end of the benchmark, the performance of MPI communications and the FACT phase on the CPU become the critical components. Our multi-threading strategy for the FACT phase helps to reduce the time spent in this regime. 

The scaling performance for this HPL implementation is observed to be over 90\% efficient when weak scaling to from a single Crusher node to 128 nodes. Full scale runs on machines such as Frontier of course require efficient scaling far beyond 128 nodes. For these large-scale runs, however, careful consideration of the performance of the MPI routines in HPL is required. It is likely that specialized communication algorithms, which optimize for the system's network topology, would be required to maintain efficient scaling, which is a topic outside the scope of this paper. Such optimizations are not present in our general implementations of these routines in \verb|rocHPL|, but the code is designed to be modular so that users can easily implement their own custom routines and further optimize for their target systems/architectures.

The steady progression of generational leaps in computational throughput on accelerated node architectures continues to pose a challenge for benchmarks such as HPL which mix compute, network bandwidth, and latency sensitive phases. As the improvement of computational throughput outpaces inter-process communication performance, the performance bottlenecks shift away from being bound by computation rate and lowers overall performance, as measured by efficiency of peak computational throughput. Future works will have to address such shifts and carefully consider how accelerators can or cannot be further leveraged in the latency- and communication-dominated tail regime of the HPL benchmark.

\section*{Acknowledgment}
The authors gratefully acknowledge the performance team at Hewlett Packard Enterprise and, in particular, Steve Whalen, Norm Troullier, and John Baron for their frequent discussions and insights. We also gratefully acknowledge the rocBLAS library team at AMD: Alex Brown, Andrew Chapman, Henry Ho, Raman Jana, Carson, Liao, Alex Liu, Wasiq Mahmood, Daine Mcniven, Koji Nakajima, Braga Natarajan, Benjamin Ulmer, Yoichi Yoshida, and Torre Zuk for all their work optimizing crucial BLAS routines on AMD Instinct GPUs.

This research used resources of the Oak Ridge Leadership Computing Facility at the Oak Ridge National Laboratory, which is supported by the Office of Science of the U.S.
Department of Energy under Contract No. DE-AC05-00OR22725.

AMD, the AMD Arrow logo, Instinct, EPYC, and combinations thereof are trademarks of Advanced Micro Devices, Inc. Other product names used in this publication are for identification purposes only and may be trademarks of their respective companies.

\printbibliography

\end{document}